\documentclass[conference]{IEEEtran}
\IEEEoverridecommandlockouts
\usepackage{cite}
\usepackage{amsmath,amssymb,amsfonts}
\usepackage{algorithmic}
\usepackage{graphicx}
\usepackage{textcomp}
\usepackage{xcolor}
\def\BibTeX{{\rm B\kern-.05em{\sc i\kern-.025em b}\kern-.08em
    T\kern-.1667em\lower.7ex\hbox{E}\kern-.125emX}}
\begin{document}

\title{A Fully Unsupervised Instance Segmentation Technique for White Blood Cell Images\\
\thanks{Identify applicable funding agency here. If none, delete this.}
}

\author{\IEEEauthorblockN{1\textsuperscript{st} Shrijeet Biswas}
\IEEEauthorblockA{\textit{A.K Chowdhury School of Information Technology} \\
\textit{University of Calcutta}\\
Calcutta, India \\
shrijeetbiswas646@gmail.com}
\and
\IEEEauthorblockN{1\textsuperscript{st} Amartya Bhattacharya}
\IEEEauthorblockA{\textit{Department of Computer Science and Engineering} \\
\textit{University of Calcutta}\\
Calcutta, India \\
amartya.bhattacharya1729@gmail.com}

}

\maketitle

\begin{abstract}
    White blood cells, also known as leukocytes are group of heterogeneously nucleated cells which act as salient immune system cells. These are originated in the bone marrow and are found in blood, plasma, and lymph tissues. Leukocytes kill the bacteria, virus and other kind of pathogens which invade human body through phagocytosis that in turn results immunity. Detection of a white blood cell count can reveal camouflaged infections and warn doctors about chronic medical conditions such as autoimmune diseases, immune deficiencies, and blood disorders. Segmentation plays an important role in identification of white blood cells (WBC) from microscopic image analysis. The goal of segmentation in a microscopic image is to divide the image into different distinct regions. In our paper, we tried to propose a novel instance segmentation method for segmenting the WBCs containing both the nucleus and the cytoplasm, from bone marrow images.
    
\end{abstract}

\begin{IEEEkeywords}
computer vision; medical imaging; unsupervised learning; healthcare
\end{IEEEkeywords}

\section{Introduction}
In medical imaging, identification of White Blood Cells (WBCs) also called leukocytes is an
important step for the diagnosis of different diseases such as leukemia, Myelodysplastic
syndromes (MDS), Acquired Immunodeficiency Syndrome (AIDS), and other immunological
syndromes. WBCs are a part of the human immune system, which exists throughout our entire
body, including bone marrow and blood. The number and types of WBCs present in a body can
serve as an indication to several health conditions and diseases, therefore; the count of
different types of WBCs called differential counting plays a major role in the determination of
the health condition of a patient \cite{1}. The traditional method of detecting WBCs involves a
trained expert who uses a microscope to select an area of interest from a bone marrow slide,
then manually detects and classifies the different White Blood Cells present in that region of
the slide. Performing all these steps manually is very difficult and tedious even for a trained
expert; therefore an unsupervised method for segmentation of WBCs from bone marrow slides
is highly desirable as it reduces the overall cost and required time to estimate the White blood
cell differential count \cite{2}.
To ease out tedious and time consuming process of manual detection of WBCs, various
automated / partially automated methods have been proposed over the past decade. Few
automated approaches which were adopted in the laboratories, used tools such as automatic
segmenting and counting machines, flow cytometry etc, to detect WBCs. To improve the quality
of detection of these tools, pattern recognition and image processing techniques can be
augmented which will enable these tools to detect and count WBCs qualitatively rather than
quantitatively \cite{3} \cite{4}. Majority of these segmentation methods are targeted towards the
application in peripheral blood rather than bone marrow. White blood cells in bone marrow are
much denser compared to those in peripheral blood. Also, immature WBCs are generally seen
only in bone marrow \cite{5}. Thus segmenting and classifying WBCs in bone morrow is more
difficult and complex compared to doing the same for peripheral blood. Therefore,
segmentation of WBCs from bone morrow slides plays a very important role in the process of
automated / partially automated differential counting of WBCs.
Segmentation of an image is the process of dividing the image into different connected regions
based on their features and properties. Segmentation techniques can be broadly classified into
two major categories, semantic segmentation and instance segmentation. In semantic
segmentation, different instances of same object are considered to be the part of the same
region of the image, whereas in instance segmentation different instances of same objects are
identified as different regions of the images. Detecting each individual WBC as a separate entity
is a challenging task of instance segmentation. Each individual slide image may contain a variety
of white blood cells which are present in different stages of maturity. As a result their nucleus
and cytoplasm may differ in shape, color, density, texture and granularities \cite{1}. Also individual
WBCs present in the slide may have overlapping cytoplasm or nucleus which makes it difficult
to determine the actual boundary between two or more WBCs. Therefore, automated

detection of each instance of WBCs present in the bone marrow slides image becomes very
challenging.
Over the years several WBC segmentation methods, both supervised and unsupervised, have
been proposed. Initially, researchers used traditional image processing techniques such as
color- space based thresholding, mathematical morphology, space scale analysis, etc to perform
the segmentation process \cite{6}\cite{7}\cite{8}\cite{9}\cite{10}\cite{11}. However, these techniques required manual
intervention. With the advent of machine learning, researchers shifted their focus from
traditional image processing methods to various machine learning models with the salient
machine learning models being \cite{12} \cite{13} \cite{14}, these models required minimal manual
intervention however, segmentation of cytoplasm and overlapping WBCs remained a problem.
With the emergence of deep learning, researchers began to employ various new deep learning
based models \cite{15}\cite{16}. These deep learning based segmentation processes required data
annotation and labeling for the purpose of generation of ground truth mask which severs as a
class label for the supervised learning process. Data annotation and generation of ground truth
masks can be very tedious and time-consuming, also it is not possible to generate a ground
truth mask for every possible type of WBCs as they might have an infinite number of variations
depending on their shape, size, density, color, texture, and granularity. All these methods
require some amount of manual intervention without which the quality output of the
segmentation process deteriorates.
In this paper, we propose a fully unsupervised technique for instance segmentation. To test the
efficacy of the system we have applied it for the segmentation and detection of WBCs form
bone marrow slide images. The approach is based on a novel combination of color space based
thresholding, K-Means Clustering followed by Watershed algorithm. The proposed method
requires no manual intervention, there is no requirement for the generation of ground truth
binary mask which is essential in majority of deep learning based segmentation methods. Also,
the proposed model is able to detect and segment out each WBC present in the slide image
including WBCs with overlapping boundaries, and output each WBC as an individual cropped
image.
\section{Related Works}
Segmentation techniques for biomedical images have been an important topic of research in the last two decades. Over the years,  approaches have been made with the help of image processing, classic machine learning, and deep learning techniques. 
Earliet works include the works of Cseke et. al 1992 \cite{10}
where authors used a method based on the Otsu Thresholding methods in order to segment the WBCs which helped the process of automatic thresholding of images for segmentation of WBCs.
Shitong et.al 2006 \cite{6} proposed a novel image processing based model for the segmentation of cells that are widely separated with the help of thresholding techniques with the subsequent application of morphological operations as well as the concept of Fuzzy cellular Network. Dorini et. al 2012 \cite{8} devised a method for nucleus as well as cytoplasm segmentation with the help of scale space analysis along with the use of morphological operations on the images. Dorini et. al, 2013 \cite{8} in his later work provided a novel method based on the Self Dual Multiscale Morphological Toggle(SMMT) algorithm along with the Watershed algorithm for segmenting nucleus and the cytoplasm separately. The authors also provided a method based on granulometric analysis combined with morphological operations for the same. 
Recent image processing based segmentations methods were implemented in the works of Safuan et. al 2017 \cite{9} where the authors provided an analysis of the performance of different color channels namely RGB,HSV and CMYK to generate the number of WBCs detected. However the research work failed to discuss about providing a method for the segmentation of cytoplasm leading to the decrease of accuracy. 
The use of various colour channels plays an important role for the segmentation job, the extensive use of which can be seen from the work done by Li et. al, 2016 \cite{11} where the authors used the combination of RGB and HSV colour channel images along with the use of dual thresholding technique in order to segment WBCs from Acute Lymphoblastic Leukemia images. \\
\\
With the advent of machine learning various algorithms were used. K Means Clustering algorithm was seen to be extensively used along with the application of various other methods on top of the algorithm. Ghane et al. 2017 \cite{13} proposed a method based on K Means Clustering algorithm. From the images of the dataset used by the authors, a clear distinction between 3 clusters namely, nucleus, cytoplasm and background, can be observed which were segmented with the help of the clustering algorithm. The segmented regions of the nucleus acted as a mask on which a modified watershed algorithm was used for the segmentation of nucleus as well as the cytoplasm region. 
The method also dealt with the problem that occurred in the cases of overlapping cells.
The authors namely implemented the method in 3 phases wherein at the first stage the author segmented the WBC region from the images, in the second stage nucleus was specifically segmented using image processing techniques like morphological operations and later in the final stage the authors solved the problem of overlapping cells which both involved detection and separation of the cells. 
Another use of K Means Clustering for segmentation of WBCs can be seen in the works of Sarrafzadeh \cite{17} where the authors used the  K Means Clustering algorithm on the LAB color space channel which involved the assumption that the elements of the bone marrow slide images should belong to 3 clusters. And each of the nucleus, cytoplasm and background can be mapped to one of them. Zheng et. al 2018 \cite{18} proposed a  self supervised approach for segmentation of WBCs. The proposed model consists of two major modules; 1) an unsupervised initial segmentation, which provides a rough segmentation result which is used to train the second part of the model, 2) an SVM classifier which helps to improve the initial segmentation results. Nasir et. al 2011\cite{19} provided a method based on the application of K Means Clustering on the hue channel image which solved the problem of segmenting the nucleus only from the images.
\\
\\
Along with the application of various machine learning algorithms, various deep learning techniques have already been applied for the task of cell segmentation. Initial stages of the work involves the application of U-Net \cite{20} where  Fully Connected Convolutional Neural Networks has been used for the segmentation task. The algorithm can applied for various supervised semantic segmentation task which involves the expensive process of labelling every images. This can also be seen in the application of Instance Segmentation techniques like Mask R CNN\cite{21}.
\\
\\
Although the methods mentioned solved the problem for segmenting the WBCs from the bone marrow images, it required manual intervention. The image processing techniques required minimal human intervention which we have tried to avoid in our method in this paper. Moreover the existing deep learning based cell segmentation techniques involve the generation of labels which can be quite an expensive task. We tried to solve this problem by providing a fully unsupervised technique for the segmentation of WBCs. 


\section{Methods and Materials}

\subsection{Data}
Data was provided by the Nightingale Hospital, Kolkata, West Bengal, India. There were 200 images shared which were bone-marrow slide images each containing White Blood Cells (WBCs). Each WBC consists of both nucleus and cytoplasm. Sample images have been shown in \ref{fig: slideimage},\ref{fig:  sampleimage2}, \ref{fig:  sampleimage3}. The images were taken under a microscope for patients having Myelodysplastic Syndrome(MDS) as well as healthy patients.

\subsection{Methods}
In our problem, the main task was to segment all the White Blood Cells(WBCs) 
 present in the bone marrow slide images. This involved developing a novel unsupervised instance segmentation technique that can automatically segment each and every WBC irrespective of its type from every bone marrow slide image. 

Our proposed algorithm mainly involves three main stages. The first stage involved segmentation of the region of interest(ROI) or the region denoting the presence of WBCs on the image, done using a semantic segmentation technique. This process helped us to create initial semantic masks of all the WBCs to be used at a later stage. 

After the image was processed in the first stage, it was passed through the second stage which involved separating the regions of cytoplasm, the background and the nucleus. For this stage K-Means clustering method was applied on the semantic masks in the first stage. As there were mainly three clusters present in each of the images, corresponding to the background, nucleus and cytoplasm respectively, the value of K was set to 3. 

And finally in order to get the instance segmentation, the watershed algorithm was used at the final stage. The watershed algorithm separated out the cells which are packed together and tough to separate using trivial image processing algorithms. Each of the  three stages has been discussed in detail, below.

\subsubsection{Stage1: Semantic Segmentation}
\begin{figure}[htp]
    \centering
    \includegraphics[width=0.45\textwidth]{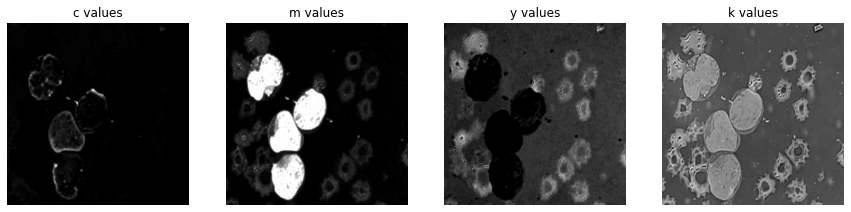}
    \caption{The C,M,Y and K channels of the original RGB image. The RGB image was changed to CMYK using the formulas mentioned. It helped in our task.}
    \label{fig:cmyk}
\end{figure}

The first stage consisted of segmenting the ROIs which represented all the WBCs along with their respective cytoplasms. The goal was accomplished
by performing operations in three steps. Initially color-space transformation was used, followed by thresholding operation. And finally after applying the morphological operations the desired result was obtained. In the first step, initially the images present in RGB(Red Green Blue) color space were transformed into CMYK(Cyan Magenta Yellow Black) color space shown in \ref{fig:cmyk}. These images in CMYK color space were used for the desired segmentation task since the contrast of the WBCs in the Y component of the image was minimum while in the M component the contrast was observed to be maximum. For transforming the images from RGB color space to CMYK color space equations 1 to 7 were used. 
\begin{equation}
R_{i}^{'} = \frac{R_{i}}{255}
\end{equation}

\begin{equation}
G_{i}^{'} = \frac{G_{i}}{255}
\end{equation}
\begin{equation}
B_{i}^{'} = \frac{B_{i}}{255}
\end{equation}
\begin{equation}
K_{i} = 1 - max(R_{i}^{'},G_{i}^{'},B_{i}^{'})
\end{equation}

\begin{equation}
C = \frac{1-R^{'} - K}{1-K}
\end{equation}
\begin{equation}
M = \frac{1-G^{'} - K}{1-K}
\end{equation}
\begin{equation}
Y = \frac{1-B^{'} - K}{1-K}
\end{equation}

where $R_{i}$ , $G_{i}$ , $B_{i}$
are the red, green and blue value of a pixel in RGB color space, and the  equivalent pixel in CMYK color space are C, M, Y, K

Histogram Equalization was applied on the images in CMYK color space, followed by contrast stretching. Since, in the Y component of the image, the contrast of the WBCs were minimum, contrast stretching made the regions, where WBCs were absent, have high contrast values thus making the regions containing WBCs distinguishable from the  rest. 

\begin{figure}[htp]
    \centering
    \includegraphics[width=0.45\textwidth]{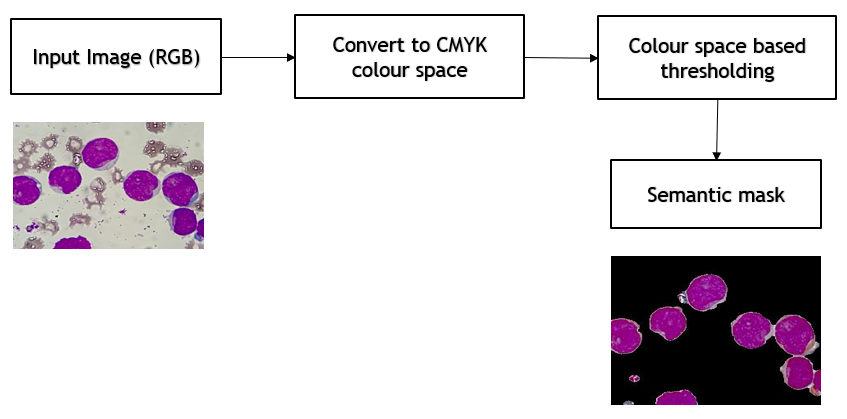}
    \caption{The various stages of operations. Initially the histogram equalized image is obtained. After that we obtained the thresholding applied and enhanced Y component image which helps in nucleus detection. After that in order to obtain the cytoplasm we used the M component. Finally we obtained the masks required.}
    \label{fig: ycomp}
\end{figure}

After the previous operations were performed, binary thresholding was applied on the resulting image followed by morphological closing which helped in obtaining the regions where WBCs were present. However in some of the cases, the previous operations led to cytoplasms getting removed from the images. In order to solve the issue, the M component was used. The M component had the maximum contrast value, in the regions where cytoplasm were present. Binary thresholding was applied on this component followed by morphological closing in order to obtain the cytoplasmic region . Now bitwise AND operation was used on these two images, one showing WBCs and the other showing the cytoplasm to get an image having both the cytoplasm as well as the nucleus . And finally the color space was changed again to RGB which has been shown in \ref{fig: ycomp}. This image was passed to stage 2. 

\subsubsection{Stage2: Application of K-Means Clustering}
\begin{figure}[htp]
    \centering
    \includegraphics[width=0.45\textwidth]{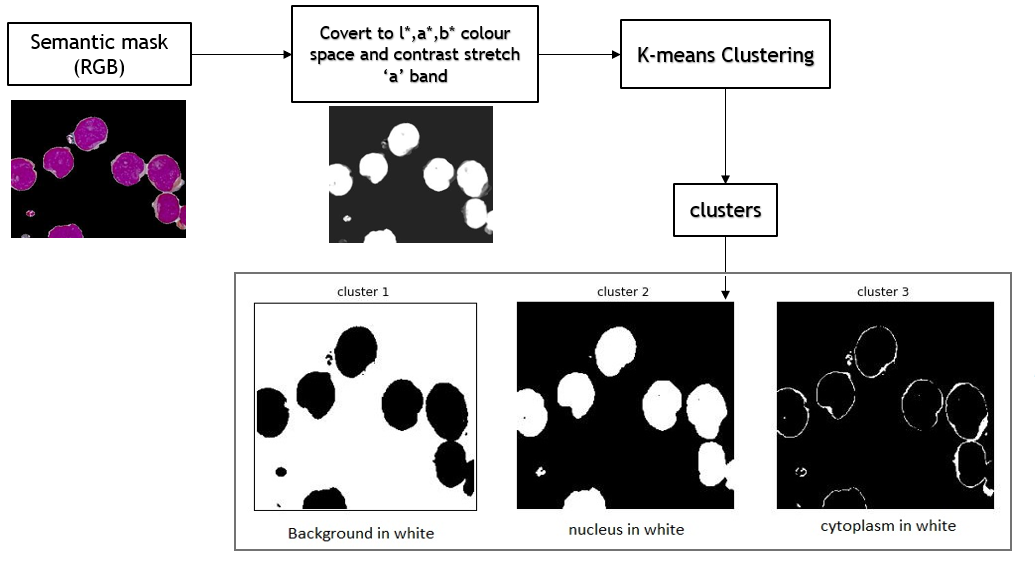}
    \caption{The results obtained after applying the K Means clustering algorithm. The K was set to 3 and each of the cluster has been shown separately.}
    \label{fig: kmeans}
\end{figure}

The semantic mask obtained after the operations performed on the first stage, was used as an input to this stage. Here the goal was to separate out each of the components present in the image. The components corresponding to the background, the nucleus and the cytoplasm were considered as three clusters. Before using the K-Means clustering algorithm on the image, the image was transformed from the RGB color space to L*A*B* color space. In this color space the intensity was represented by the L* color channel. This image can clearly discriminate the nuclei from the rest of the image since the lightness value of the nuclei is more than the other regions. The “a” component of the image was selected and contrast stretching was applied to create more difference in pixel values between the region corresponding to the nucleus and the region corresponding to the cytoplasm. The contrast stretched image was used as the input to the K-Means algorithm and K was set to 3 as discussed before. The output of the image after using the K-Means clustering algorithm has been shown in \ref{fig: kmeans}. However the K-Means algorithm, although can segment the three regions separately, it doesn’t classify and has an unique class no. for each of the regions containing nucleus, cytoplasm and background. This might lead to inconsistencies which can prove to be problematic at the later stage of the algorithm. This issue was solved using the concept of Intersection over Union(IoU) whose equation has been given in the equation below. 

IoU = Intersection of two areas/ Union of the two areas

Initially a basic nucleus mask was generated by converting the image into a grayscale image and binary thresholding was applied. After that the IoU was calculated using the mask corresponding to the region containing nucleus and 
each of the three clusters. The cluster which had the maximum and  IoU value was considered as the region of nucleus and the region of background respectively. While the other region was considered to be the region containing the cytoplasm. In this way the issue of inconsistency between the clusters was solved. 

\subsubsection{Stage 3: Traditional Watershed Algorithm}
\begin{figure}[htp]
    \centering
    \includegraphics[width=0.45\textwidth]{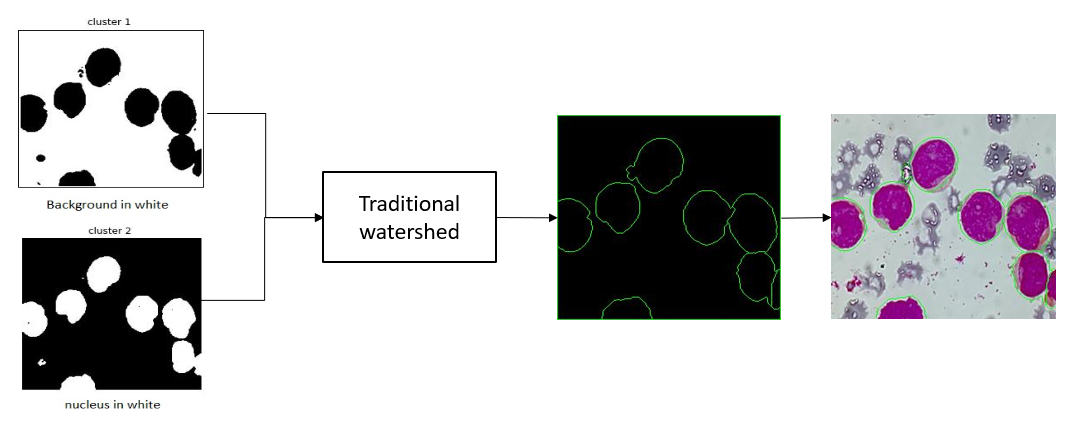}
    \caption{The centers and the region that was filled using the watershed algorithm has been shown.}
    \label{fig: watershed1}
\end{figure}

This was used in the final stage of our proposed methodology and it helped us make instance segmentation, separating out each of the WBCs with a clear boundary. 
For this stage, the images having the nucleus cluster and the background cluster, obtained from the previous stage, were used as an input. Watershed algorithm requires the presence of at least one marker or seed values, which are obtained by applying distance transform on the cluster containing the nucleus, inside each of the objects present in the image, including the background as a separate object. Once the seed values are generated, each object present in the image is marked. These seeds can then be grown using a morphological watershed transformation. During this process the seeds will touch each other in the cases where the WBCs are in direct contact with each other. Whenever this happens, the region where there is direct contact between the two seeds is considered as the boundary between the two or more seeds or WBCs.  
The results of this process has been shown in \ref{fig: watershed1}

\subsection{Experiments}
After the acquisition of the data was successful, the above-mentioned technique was applied to the given images, and the outcomes were observed. 
Each of the images was taken one at a time and the method was applied to each of the images. After the proposed method was applied, images were obtained showing the boundaries of each of the cells separately. This process was repeated for all of the images in our dataset. And finally, as a measure of the performance, all the IOUs were calculated.

\subsection{Results}
\begin{figure}[htp]
    \centering
    \includegraphics[width=0.45\textwidth]{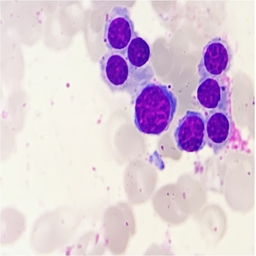}
    \caption{This shows a sample of a bone marrow slide image. It contains normal as well as dysplastic cells.}
    \label{fig: slideimage}
\end{figure}

\begin{figure}[htp]
    \centering
    \includegraphics[width=0.45\textwidth]{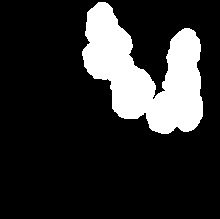}
    \caption{The above image shows the semantic segmentation masks generated. This proves crucial for the next stages involved. Here we can clearly differentiate the background with the cells.}
    \label{fig: semanticcell}
\end{figure}

\begin{figure}[htp]
    \centering
    \includegraphics[width=0.45\textwidth]{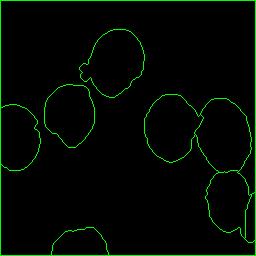}
    \caption{This shows the contours of each and every segmented cell present in the bone-marrow slide image.}
    \label{fig: contour}
\end{figure}

\begin{figure}[htp]
    \centering
    \includegraphics[width=0.45\textwidth]{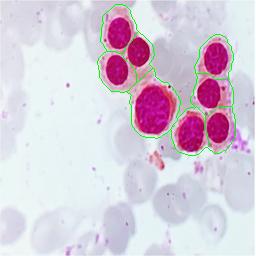}
    \caption{The result obtained after performing the techniques mentioned in our proposed framework.}
    \label{fig: finalresult}
\end{figure}

\begin{figure}[htp]
    \centering
    \includegraphics[width=0.45\textwidth]{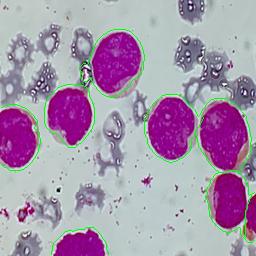}
    \caption{The result obtained after performing the techniques mentioned in our proposed framework.}
    \label{fig: finalresult1}
\end{figure}

\begin{figure}[htp]
    \centering
    \includegraphics[width=0.45\textwidth]{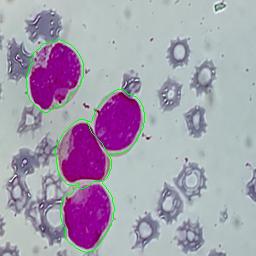}
    \caption{The result obtained after performing the techniques mentioned in our proposed framework.}
    \label{fig: finalresult2}
\end{figure}

\begin{figure}[htp]
    \centering
    \includegraphics[width=0.45\textwidth]{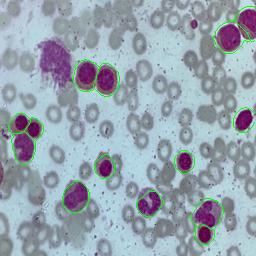}
    \caption{The result obtained after performing the techniques mentioned in our proposed framework.}
    \label{fig: finalresult3}
\end{figure}

The proposed framework was used in order to segment the cells from 200 bone marrow slide images. Sample slide image has been attached below \ref{fig: slideimage}. As mentioned we applied the framework in three different stages. First, we applied the semantic segmentation technique, that helped us to obtain the corresponding segmentation mask for the bone-marrow slide image. The sample results has been shown in \ref{fig: semanticcell}. After obtaining the semantic segmentation masks of the slide image, we apply the K-Means Clustering method as discussed above and also the Watershed technique. The result obtained after applying these methods has been shown in \ref{fig: contour}. We finally overlay the contours with the original image to obtain the final results i.e segmenation of each and every cell present in the bone-marrow slide image as showin in \ref{fig: finalresult}. Similarly \ref{fig: finalresult1} \ref{fig: finalresult2} \ref{fig: finalresult3} shows sample results obtained after applying our proposed framework on different slide images. 

After calculating the Intersection over Union(IOU) with the manually masked images, the IOU was found to have an average of 0.85 or 85\%, calculated over 200 images. 



\subsection{Conclusion}
Instance Segmentation of WBCs from bone marrow images is a challenging task when the WBCs are densely populated in a single bone marrow slide image. Using supervised algorithms to solve this problem, generally require a large number of image samples along with their labels. Generating the correct labels manually becomes a tough task and any mistake in the label can lead to an incorrect solution. Here in our paper we present a multi stage unsupervised instance segmentation model which is capable of segmenting every WBCs containing both the nucleus and the cytoplasm from the bone marrow slide images and separating each of the cells with a boundary. The solution involves three separate steps of creating a mask, using K-Means Clustering algorithm and then using Watershed algorithm to segment each of the WBCs. 
Our solution performs better than the existing unsupervised as well as the supervised methods that have been proposed earlier and also requires minimal number of parameters thus computationally efficient.


\begin{thebibliography}{1}
\bibitem{1} Alberts, Bruce, et al. ;Cell junctions.; Molecular Biology of the Cell. 4th edition. Garland Science, 2002.

\bibitem{2}Theera-Umpon, Nipon, and Sompong Dhompongsa. ;Morphological granulometric features
of nucleus in automatic bone marrow white blood cell classification.;IEEE Transactions on
Information Technology in Biomedicine11.3 (2007): 353-359.
\bibitem{3} Clark, Vivian L., and James A. Kruse.;Clinical methods: the history, physical, and laboratory
examinations.;Jama264.21 (1990): 2808-2809.
\bibitem{4} Al-Dulaimi, Khamael Abbas Khudhair, et al. ;Classification of white blood cell types from
microscope images: Techniques and challenges.;Microscopy science: Last approaches on
educational programs and applied research (Microscopy Book Series, 8)(2018): 17-25.
\bibitem{5} Minnich, Virginia.Immature Cells in the Granulocytic, Monocytic, and Lymphocytic Series.
American Society of Clinical Pathologists, 1982.
\bibitem{6} Shitong, Wang, and Wang Min.;A new detection algorithm (NDA) based on fuzzy cellular
neural networks for white blood cell detection..;IEEE Transactions on information technology in
biomedicine 10.1 (2006): 5-10.
\bibitem{7}Dorini, Leyza Baldo, Rodrigo Minetto, and Neucimar Jerônimo Leite. ;White blood cell
segmentation using morphological operators and scale-space analysis.; XX Brazilian Symposium
on Computer Graphics and Image Processing (SIBGRAPI 2007). IEEE, 2007.
\bibitem{8}
Dorini, Leyza Baldo, Rodrigo Minetto, and Neucimar Jeronimo Leite. ;Semiautomatic white
blood cell segmentation based on multiscale analysis..; IEEE journal of biomedical and health
informatics 17.1 (2012): 250-256.
\bibitem{9}Safuan, Syadia Nabilah Mohd, et al. .;White blood cell counting analysis of blood smear
images using various segmentation strategies..; AIP Conference Proceedings. Vol. 1883. No. 1.
AIP Publishing LLC, 2017.
\bibitem{10}Cseke, Istvan. .;A fast segmentation scheme for white blood cell images..; Proceedings.,
11th IAPR International Conference on Pattern Recognition. Vol. III. Conference C: Image,
Speech and Signal Analysis,. IEEE, 1992.
\bibitem{11}Li, Yan, et al. ;Segmentation of white blood cell from acute lymphoblastic leukemia images
using dual-threshold method..; Computational and mathematical methods in medicine 2016
(2016).
\bibitem{12}Zhang, Congcong, et al. .;White blood cell segmentation by color-space-based k-means
clustering.; Sensors 14.9 (2014): 16128-16147.
\bibitem{13}Ghane, Narjes, et al. .;Segmentation of white blood cells from microscopic images using a
novel combination of K-means clustering and modified watershed algorithm.; Journal of
medical signals and sensors 7.2 (2017): 92.
\bibitem{14}Zheng, Xin, et al. .;Fast and robust segmentation of white blood cell images by self-
supervised learning.; Micron 107 (2018): 55-71.
\bibitem{15}Wang, Qiwei, et al.;Deep learning approach to peripheral leukocyte recognition.; PloS
one 14.6 (2019): e0218808.
\bibitem{16}Dhieb, Najmeddine, et al .;An automated blood cells counting and classification framework
using mask R-CNN deep learning model..; 2019 31st International Conference on
Microelectronics (ICM). IEEE, 2019.
\bibitem{17}Sarrafzadeh, Omid, and Alireza Mehri Dehnavi. "Nucleus and cytoplasm segmentation in microscopic images using K-means clustering and region growing." Advanced biomedical research 4 (2015).
\bibitem{18}Sarrafzadeh, Omid, and Alireza Mehri Dehnavi. "Nucleus and cytoplasm segmentation in microscopic images using K-means clustering and region growing." Advanced biomedical research 4 (2015).
\bibitem{19}Nasir, AS Abdul, M. Y. Mashor, and H. Rosline. "Unsupervised colour segmentation of white blood cell for acute leukaemia images." 2011 IEEE International Conference on Imaging Systems and Techniques. IEEE, 2011.
\bibitem{20}Ronneberger, Olaf, Philipp Fischer, and Thomas Brox. "U-net: Convolutional networks for biomedical image segmentation." International Conference on Medical image computing and computer-assisted intervention. Springer, Cham, 2015.
\bibitem{21}He, Kaiming, et al. "Mask r-cnn." Proceedings of the IEEE international conference on computer vision. 2017.

\end{thebibliography}
\end{document}